\documentclass[showpacs,preprintnumbers,amsmath,amssymb]{revtex4}
\usepackage{graphicx}
\usepackage{dcolumn}
\makeatletter
\input{epsf}

\begin{document}
%\preprint{APS/123-QED}
\title{Magnus Force on Quantum Hall Skyrmions and Vortices}

\author{S. Dhar}
 \email{sarmishtha_r@isical.ac.in}
\author{B. Basu}
\email{banasri@isical.ac.in}
\author{P. Bandyopadhyay}
 \email{pratul@isical.ac.in}
\affiliation{Physics and Applied Mathematics Unit\\
 Indian Statistical Institute\\
 Calcutta-700108 }
%\date{\today}

\begin{abstract}
We have discussed here the Magnus force acting on the  vortices
and skyrmions in the quantum Hall systems. We have found that it
is generated by the chirality of the system which is associated
with the Berry phase and is same for both the cases.
\end{abstract}
\pacs{11.15.-q, 12.39.Dc, 73.43.-f}
\maketitle

%\section{Introduction}
In recent times, the topological excitations near the filling
factor $\nu= 1$ in quantum Hall effect have drawn much attention
\cite{1,2,3}. There are two kinds of topological excitations in
single layer quantum Hall systems. When the system is fully
polarized the relevant charged quasiparticles are {\it topological
vortices}. At $\nu=\frac{1}{2n+1}$ the charge and spin of such a
vortex is $\frac{e}{2n+1}$ and $\frac{1}{2(2n+1)}$ respectively.
In a remarkable paper Sondhi et. al. \cite{1} have argued that for
a weak Zeeman coupling the lowest energy charged   quasiparticle
is a {\it topological soliton} or {\it skyrmion}. The charge is
still $\frac{e}{2n+1}$ but the total spin can be substantially
larger. This large spin with moderate charge explains the observed
depolarization when the filling factor slightly deviates from
$\nu=\frac{1}{2n+1}$.  Indeed, the difference between a vortex and
a skyrmion is that, a vortex is fully polarized while inside the
skyrmion core there is some number of electrons with reversed
spins. It may be noted that there is an interplay of two factors,
namely the Zeeman and the Coulomb energy. Vortex solutions exists
for any value of the Zeeman coupling and is well known from the
studies on the vortices in the Ginzburg-Landau model of the fully
polarized quantum Hall effect \cite{4}. For weak Zeeman coupling
the relevant quasiparticles are the skyrmions. In this letter, we
have  studied  the Magnus force acting on these skyrmions and
vortices from their topological properties when it is studied in
terms of $3+1$ dimensional nonlinear sigma model and found that
they are the same.

 To investigate the dynamics of vortices and skyrmions we begin with the
Landau-Ginzburg theory of the Hall effect introduced by Zhang,
Hansson and Kivelson \cite{5}  and modified by Lee and Kane
\cite{6} to incorporate the effect of the spin. For spin $1/2$
particles if we set $\hbar=e=c=1$, the Lagrangian can be written
as

\begin{eqnarray}
L=&&\overline{\Psi}_\sigma[\partial_0-i(a_0+ A_0)]\Psi_\sigma-\frac{1}{2m^*}|[\partial_{i}-i(a_{i}+A_{i})]\Psi_\sigma|^2\nonumber\\
&&-\frac{\lambda}{2}(\overline{\Psi}_\sigma\Psi_\sigma-\rho_0)^2+\frac{1}{4\theta}\epsilon^{\mu\nu\alpha}a_{\mu}\partial_{\nu}a_{\alpha}
\end{eqnarray}
Here $\Psi_\sigma$ is a two-component Bose field with effective
mass $m^*$ and short-range repulsive interaction $\lambda$, which
couples to the external and statistical gauge fields $A_i$ and
$a_i(i=0,1,2)$. Here $\rho_0$ is the uniform density
$|\phi_1|^2+|\phi_2|^2$ at filling factor $\nu=\frac{1}{2n+1}$. In
order to separate the charge and spin degree of freedom, we
explicitly separate the magnitude and $U(1)$ and $SU(2)$ phases of
$\Psi_\sigma:~\Psi_\sigma\rightarrow \sqrt{\rho} \phi z_\sigma$
with $\overline{\phi}\phi=\overline{z_\sigma}z_\sigma=1$. By
direct substitution and keeping the leading-order gradient terms
we obtain
\begin{eqnarray}
L=&&\rho[\overline{\phi}\partial_0
\phi+\overline{z_\sigma}\partial_0 z_\sigma-i(a_0+ A_0)
]-\frac{\rho}{2m^*}|\overline{\phi}\partial_i
\phi+\overline{z}_\sigma\partial_i
z_\sigma-i(a_i+A_i)|^2\nonumber\\
&&-\frac{\rho}{2m^*}[|\partial_i
z_\sigma|^2+(\overline{z}_\sigma\partial_i
z_\sigma)^2]-\frac{\lambda}{2}(\rho-\rho_0)^2+\frac{1}{4\theta}\epsilon^{\mu\nu\alpha}a_{\mu}\partial_{\nu}a_{\alpha}
\end{eqnarray}
We now note that we have the identity
\begin{equation}
\frac{\rho}{2m^*}[|\partial_i
z_\sigma|^2+(\overline{z}_\sigma\partial_i
z_\sigma)^2]=\frac{\rho}{8m^*}~(\nabla{\bf n})^2
\end{equation}
where $\nabla{\bf n}=(\partial_1{\bf n},\partial_2{\bf n})$ with
${\bf n}=\bar{z}_\sigma \sigma_{\alpha\beta} z_\sigma$,
$\sigma_{\alpha\beta}$ being Pauli matrices. We now introduce
Hubbard-Stratonovich fields ${\bf J}$ to decouple the second term
of eqn.(2) to obtain
\begin{eqnarray}
L=&&i \rho[\overline{\phi}\partial_0
\phi+\overline{z}_\sigma\partial_0 z_\sigma-i(a_0+
A_0)]+i[\overline{\phi}\partial_i
\phi+\overline{z}_\sigma\partial_i z_\sigma-i(a_i+ A_i)]J_i+\frac{m^*}{2\rho}|{\bf J}|^2\nonumber\\
&&-\frac{\rho}{8m^*}~(\nabla{\bf
n})^2-\frac{\lambda}{2}(J_0-\rho_0)^2+\frac{1}{4\theta}\epsilon^{\mu\nu\alpha}a_{\mu}\partial_{\nu}a_{\alpha}
\end{eqnarray}
where ${\bf J}$ is the three vector $\rho=(J_0,J_1,J_2)$. After
integrating out the longitudinal fluctuations in $\phi$, we have
the conserved current relation $\partial_\mu J_\mu =0$, which can
be satisfied by taking $J_\mu$ as the curl of a three dimensional
vector field. We can now set $J_{\mu}^{[0]}=(\rho_0,0,0)$ equal to
$\epsilon_{\mu\nu\lambda}\partial_{\nu}{\cal{A}}^{[0]}_{\lambda}$
and

\begin{equation}
J_{\mu}-J_{\mu}^{[0]}=\epsilon_{\mu\nu\lambda}\partial_{\nu}\mathcal{A}_{\lambda}
\end{equation}

Now, following Stone \cite{7}, we integrate out the Chern-Simons
field $a_\mu$ and using the relation $2\theta\rho_0=eB_z$, we can
write

\begin{equation}
\begin{array}{lcl}
L &=&\displaystyle{2\pi
[{\cal{J}}^{V}_{\mu}({\cal{A}}_{\mu}+{\cal{A}}^{[0]}_{\mu})+{\cal{J}}^{S}_{\mu}({\cal{A}}_{\mu}+{\cal{A}}^{[0]}_{\mu})]
-\theta {|{\mathbf{J}}|^{\mu}} {\cal{A}}_{\mu}}\\
&&\displaystyle{+\frac{m^{\ast}}{2\rho} {\bf{J}}^2
-\frac{\rho}{8m^{\ast}}(\nabla{\bf{n}})^2
-\frac{\lambda}{2}(J_0-J_0^{[0]})^2}\\
\end{array}
\end{equation}
where
\begin{eqnarray}
{\mathcal{J}}^{V}_\mu=&& {\frac{1}{2{\pi} i}}
\epsilon_{\mu\nu\lambda}\partial_{\nu} \bar{\phi}\partial_\lambda
\phi\\
{\mathcal{J}}^{S}_\mu=&&{\frac{1}{2{\pi} i}}
\epsilon_{\mu\nu\lambda}\partial_{\nu}\bar{z}_{\sigma}\partial_{\lambda}z_{\sigma}
\end{eqnarray}
are the skyrmion and vortex  three currents, respectively. It is
observed that the skyrmion current can be written in the familiar
form
\begin{equation}
{\mathcal{J}}^{S}_\mu= {\frac{1}{4{\pi} }}
\epsilon_{\mu\nu\lambda}{\bf n}.(\partial_\nu {\bf n} \times
\partial_\lambda {\bf n})
\end{equation}

Now setting $\rho=\rho_0$ in the kinetic energy term and adjusting
the units of length and time such that
$c=\sqrt{\lambda\rho_0/m^*}$, the velocity of density wave in the
absence of the magnetic field becomes unity and defining the field
strength tensor
$\cal{F}_{\mu\nu}=\partial_{\mu}\cal{A}_{\nu}-\partial_{\nu}\cal{A}_{\mu}$,~
we can write

\begin{equation}
\begin{array}{lcl}
L &=&\displaystyle{2\pi
[{\cal{J}}^S_{\mu}({\cal{A}}_{\mu}+{\cal{A}}^{[0]}_{\mu})+{\cal{J}}^{V}_{\mu}({\cal{A}}_{\mu}+{\cal{A}}^{[0]}_{\mu})]-\frac{1}{2}\theta\epsilon_{\mu\nu\sigma}{\cal{A}}_{\mu}{\cal{F}}_{\nu\sigma}}\\
&&\displaystyle{-\frac{\lambda}{4}{\cal{F}}_{\mu\nu}{\cal{F}}^{\mu\nu}-\frac{1}{8\lambda}(\nabla{\bf{n}})^2}\\
\end{array}
\end{equation}

It is observed that here ${\cal A}_\mu$ is a topologically massive
gauge field and ${\cal A}^{[0]}_\mu$ just represents the
background field.

To study the Magnus force on these vortices and skyrmions, we now
take resort to the spherical geometry. In a recent paper \cite{8},
we have studied quantum Hall skyrmions in terms of the $3+1$
dimensional nonlinear sigma model where we have taken a system of
$2D$ electron gas residing on the surface of a $3D$ sphere with a
monopole at the centre. We note that taking the spin variable
${\bf{z}}=U{\bf{z_0}}$ where ${\bf{z}}_0=\left(\begin{array}{c}
  1 \\
  0
\end{array}\right)$
and $U\in SU(2)$, we may write the nonlinear sigma model
Lagrangian in terms of $U$. The $3+1$ dimensional generalization
of the skyrmion current can now be defined as
\begin{equation}
{\mathcal{J}}^S_{\mu}~=~\frac{1}{24 \pi^2}~
\epsilon_{\mu\nu\lambda\sigma}
Tr[(U^{-1}\partial_{\nu}U)(U^{-1}\partial_{\lambda}U)
(U^{-1}\partial_{\sigma}U)]
\end{equation}
where it is associated with an $O(4)$ nonlinear sigma model with
$U$ defined as
\begin{equation}
U=\pi_0+i
\overrightarrow{\pi}.\overrightarrow{\sigma},~~~~~~~~U\in SU(2)
\end{equation}
where $\overrightarrow{\sigma}$ are Pauli matrices and
$\overrightarrow{\pi}$ are chiral boson fields, satisfying the
constraint
\begin{equation}
\pi^2_0+\overrightarrow{\pi}^2=1
\end{equation}

The $3+1$ dimensional generalization of the vortex current can be
written as
\begin{equation}
{\mathcal{J}}^{V}_\mu=
\epsilon_{\mu\nu\lambda\sigma}\partial_{\nu}
{\phi}\partial_\lambda \phi\partial_\sigma \phi
\end{equation}
In a pure gauge, we can take a gauge field $B_\mu$ such that
\begin{equation}
B_\mu=\partial_\mu \phi
\end{equation}

Now noting that vortex -antivortex pair can be taken as an $SU(2)$
doublet, we may consider $B_\mu$ as an $SU(2)$ gauge field and
write
\begin{equation}
B_\mu=\tilde{U}^{-1}\partial_\mu \tilde{U},~~~~~~~~~\tilde{U}\in
SU(2)
\end{equation}
In view of this, we can write the vortex current taking into
account proper normalization
\begin{equation}
{\mathcal{J}}^V_{\mu}~=~\frac{1}{24 \pi^2}~
\epsilon_{\mu\nu\lambda\sigma}
Tr[(\tilde{U}^{-1}\partial_{\nu}\tilde{U})(\tilde{U}^{-1}\partial_{\lambda}\tilde{U})
(\tilde{U}^{-1}\partial_{\sigma}\tilde{U})]
\end{equation}

Thus comparing eqn.(11) and eqn.(17) we note that the skyrmion
current and vortex current can be written in a similar form which
is also shown by Duan et. al. \cite{15}.

In $3+1$ dimensions we can generalize the Lagrangian (10) with
non-Abelian gauge field ${\cal A}_\mu[{\cal A}^{[0]}_\mu] \in
SU(2)$ and the $\theta$-term in the form
\begin{equation}
\begin{array}{lcl}
L &=&\displaystyle{2\pi
[{\cal{J}}^S_{\mu}({\cal{A}}_{\mu}+{\cal{A}}^{[0]}_{\mu})+{\cal{J}}^{V}_{\mu}({\cal{A}}_{\mu}+{\cal{A}}^{[0]}_{\mu})]-~\frac{M^2}{16}~Tr(\partial_{\mu}
U^{\dag}\partial_\nu U)}\\
&&\displaystyle{-\frac{\theta}{16\pi^2}~Tr~
{^{*}{\mathcal{F}}_{\mu\nu}}{\mathcal{F}}_{\mu\nu}-\frac{1}{4}~Tr~{\mathcal{F}}_{\mu\nu}{\mathcal{F}}^{\mu\nu}}\\
\end{array}
\end{equation}
Here $\theta=g/c^2$ with $g=\nu e^2/h$ as Hall conductivity and
$~^{*}{\mathcal{F}}_{\mu \nu}$   is a Hodge dual given by
\begin{equation}
^{*}{\mathcal{F}}_{\mu\nu}
=\frac{1}{2}~\epsilon_{\mu\nu\lambda\sigma}{\cal
F}_{\lambda\sigma}
\end{equation}

The third term in eqn(18) is the $\theta$-term which is the $3+1$
dimensional relative to the $2+1$ dimensional Chern-Simons term.
It is noted that the $\theta$-term is related to chiral anomaly
and Berry phase \cite{10}. Indeed, it is a four divergence and we
can write

\begin{equation}
-{\frac{1}{16 \pi^2}}~ Tr~ ^{*} {\mathcal{F}}_{\mu\nu}
{\mathcal{F}}_{\mu\nu}=\partial_\mu \Omega_\mu
\end{equation}
where $\Omega_\mu$ is the Chern-Simons characteristic class given
by
\begin{equation}
\Omega_{\mu} = -{\frac{1}{16 \pi^2}} \epsilon_{\mu \nu \lambda
\sigma}
 Tr({\cal{A}}_{\nu} {\cal{F}}_{\lambda \sigma} + {\frac{2}{3}} {\cal{A}}_{\nu}{\cal{A}}_{\lambda}{\cal{A}}_{\sigma})
\end{equation}

When a fermionic chiral current interacts with a gauge field, we
may define
\begin{equation}
\tilde{J}^{5}_\mu={J}^{5}_\mu+2~\Omega_{\mu}
\end{equation}
where ${J}^{5}_\mu$ is the axial vector current so that
$\bar{\Psi}\gamma_\mu \gamma_5 \Psi$, we have $\partial_\mu
\tilde{J}^{5}_\mu=0$ whereas $\partial_\mu{J}^{5}_\mu\neq 0$.
Indeed, we have the chiral anomaly given by
\begin{equation}
\partial_\mu{J}^{5}_\mu=-2~\partial_\mu \Omega_\mu={\frac{1}{8 \pi^2}}~ Tr~ ^{*} {\mathcal{F}}_{\mu\nu}
{\mathcal{F}}_{\mu\nu}
\end{equation}

The Pontryagin index is given by
\begin{eqnarray}
 q~=&&~2\mu\nonumber\\
 =&&-{\frac{1}{16 \pi^2}}~\int Tr ~^{*} {\mathcal{F}}_{\mu\nu} {\mathcal{F}}_{\mu\nu} d^4 x\nonumber\\
=&&\int d^4 x \partial_\mu
\Omega_\mu\nonumber\\
=&&-\frac{1}{2}~\int~\partial_{\mu}J^5_{\mu} d^4 x
\end{eqnarray}
Here $\mu$ represents a magnetic charge. The Berry phase
$e^{i\Phi_B}$ is associated with the chiral anomaly \cite{9,10}
through the relation

\begin{equation}
\Phi_B=2 \pi \mu
\end{equation}
 In Euclidian
space-time if we demand ${\cal F}_{\mu\nu}=0$, then the gauge
potential tends to a pure gauge in the limiting case towards the
boundary i.e. we can take
\begin{equation} {\cal{A}}_{\mu} = U^{-1}
\partial_{\mu} U,~~~~~~~~~~~~~~~~U\in SU(2)
\end{equation}
This gives
\begin{equation}
\Omega_\mu=-\frac{1}{24 \pi^2}~ \epsilon_{\mu\nu\lambda\sigma}
Tr[(U^{-1}\partial_{\nu}U)(U^{-1}\partial_{\lambda}U)
(U^{-1}\partial_{\sigma}U)]
\end{equation}

Comparing this with eqn(11) we note that $\Omega_\mu$ can be
related to the skyrmion current. It is noted that on the boundary
$S^3$ where ${\cal F}_{\mu\nu}=0$, we have $\partial_\mu
\Omega_\mu=0$ as is evident from eqn.(23). However, inside the
volume $V^4$ where ${\cal F}_{\mu\nu}\neq 0$, $\Omega_\mu$ is
associated with Berry phase as follows from eqn.(24). On the
boundary $S^3$ of a volume $V^4$, the term ${\cal F}_{\mu\nu}{\cal
F}^{\mu\nu}$ in the Lagrangian (18) gives rise to the skyrme term
$[\partial_\mu U U^{-1},\partial_\nu U U^{-1}]^2$ in this limiting
case. This ensures the stability of the skyrmion so that it does
not shrink to zero size.
 Here also ${\cal A}_\mu$
appears as a topologically massive gauge field as it has been
shown elsewhere that when a gauge field interacts with a chiral
current it acquires mass topologically through chiral anomaly
\cite{11}.

If we take the compactified $3$-sphere we have the winding number
associated with the homotopy $\Pi_3(S^3)=Z$
\begin{equation}
q=2\mu=\frac{1}{24 \pi^2}~ \int_{S^3} dS_\mu
\epsilon_{\mu\nu\lambda\sigma}
Tr[(U^{-1}\partial_{\nu}U)(U^{-1}\partial_{\lambda}U)
(U^{-1}\partial_{\sigma}U)]
\end{equation}

This effectively represents the geometric phase associated with
the $\theta$-term in the nonlinear sigma model action.

The configuration of a skyrmion is such that spins wrap an unit
sphere with a Dirac flux quanta within it and the resultant spin
arising out of spin reversals will give rise to a specific
chirality. When a skyrmion of charge $\alpha$
$(\alpha=\frac{e}{2n+1})$ moves around a closed path, the Berry
phase is given by $2\pi \alpha N$ where $N$ is the number of
skyrmions enclosed by the path as is evident from eqn.(25). The
Magnus force \cite{13} is now given by
\begin{equation}
F=2 \pi \alpha N \hat{c} \times {\bf V}
\end{equation}
where $\hat{c}$ represents the axis of resultant chirality and
${\bf V}$ is the velocity of the skyrmion with respect to the
quantum Hall fluid.

%discussion

When the Zeeman energy is large, i.e. the vortices are the
dominant excitations in the system, these vortices will also
experience a similar Magnus force. Since the two currents are of
similar form, the vortex current can also be associated with the
chiral anomaly giving rise to the Berry phase. We know that
topologically a vortex is equivalent to a magnetic flux, so when a
vortex moves in a closed path, the Berry phase is given by $2 \pi
\alpha N$ where $\alpha=\frac{e}{2n+1}$ is the charge of the
vortex and $N$ is the number of vortices enclosed by the loop. The
Magnus force is given by the vector product of the vorticity and
the velocity relative to the quantum Hall fluid.
\begin{equation}
F=\pm 2 \pi \alpha N \hat{z} \times {\bf V}_{vortex}
\end{equation}
where $\pm$ corresponds to a vortex parallel (antiparallel) to the
$z$-direction and ${\bf V}_{vortex}$ is the vortex velocity. Thus
a comparison between eqn.(29) and eqn.(30) suggests that the
Magnus force experienced by a vortex and a skyrmion in a quantum
Hall fluid is the same. This is consistent with the result
obtained by Dziarmaga \cite{12}.

This analysis supports  the Ao-Thouless theory \cite{13} of Magnus
force which associates the Magnus force with the Berry phase. In
this scenario, the skyrmion current as well as the vortex current
effectively represents a chiral current which is associated with
the chiral anomaly when it interacts with a topologically massive
gauge field. The Magnus  force is generated by the background
field associated with the chirality of the system. Finally, we may
mention that in a recent paper \cite{14} we have analyzed the
Magnus force acting on vortices in high temperature
superconductivity and is also found to be generated by the
background field associated with the chirality of the system. We
observe here that if we study the Magnus force from the
topological properties of the excitations concerned the Magnus
force acting on vortices and skyrmions in quantum Hall fluid as
well as on vortices in high T$_c$ superconductivity is of similar
origin.

%\section{Discussion}

%{\bf References}

\end{document}